\documentclass[prd,aps,12pt]{revtex4}
\usepackage{epsfig}
\begin{document}

\title{Spectrum of quenched twisted mass lattice QCD at maximal twist}

\author{Abdou M. Abdel-Rehim and Randy Lewis}

\affiliation{Department of Physics, University of Regina, Regina, SK,
             Canada, S4S 0A2}

\author{R. M. Woloshyn}

\affiliation{TRIUMF, 4004 Wesbrook Mall, Vancouver, BC, Canada, V6T 2A3}

\begin{abstract}
Hadron masses are computed from quenched twisted mass lattice QCD for a
degenerate doublet of up and down quarks with the
twist angle set to $\pi/2$, since this maximally twisted theory is expected
to be free of linear discretization errors.  Two separate definitions of
the twist angle are used, and the hadron masses for these two cases are
compared.  The flavor breaking, that can arise due to twisting, is discussed
in the context of mass splittings within the $\Delta(1232)$ multiplet.
\end{abstract}

\maketitle

\section{Introduction}\label{sec:intro}

Twisted mass lattice quantum chromodynamics (tmLQCD) is an
extension of the familiar Wilson lattice action \cite{tmLQCD0, tmLQCD, early}.  
TmLQCD is essentially obtained by introducing a chiral-flavor rotation in the 
mass term of the Wilson action for a quark doublet
(see Sec.~\ref{sec:action}), and it offers
a number of significant advantages.  TmLQCD does not suffer from the
``exceptional configurations'' which arise in quenched and partially quenched
Wilson simulations \cite{Bardeen} due to
the existence of unphysical zero modes.  
In tmLQCD, $O(a)$ improvement (where $a$ is the lattice spacing)
can be implemented in a convenient manner. In particular at maximal twist
$O(a)$ improvement is often automatic \cite{FreRos}, that is, without 
the addition of a clover term \cite{clover}.
The chiral nature of the rotation leads to parity mixings, but at maximal
twist the mixings become mere operator interchanges.
In addition, dynamical tmLQCD simulations \cite{dynamical,dynamical2}
are dramatically less expensive
than dynamical chiral fermion simulations, and are claimed to be comparable
to the cost of dynamical staggered fermions \cite{Kenn}.

Nevertheless there are issues \cite{dynamical,dynamical2,SharpeWu1,AokiBar}
that remain to be understood.
Of particular concern in this work is the definition of maximal twist.
The original suggestion of Frezzotti and Rossi \cite{FreRos} was to 
obtain maximal twist by using the critical hopping parameter ($\kappa_{cW}$)
from the Wilson action. This procedure (which we refer to as the Wilson 
$\kappa_{cW}$ definition)
was used in previously published computations \cite{Jansen,GoingChiral,pionff}.
However, recent theoretical studies \cite{AokiBar,SharpeWu2} have suggested 
that this is generally not precise enough to maintain tmLQCD's $O(a)$ 
improvement for small quark masses (for another point of view see 
Ref. \cite{FreLat04}), and specific alternatives were presented.

In this work, the effect of using different definitions of maximal twist is 
studied by performing two separate simulations of the quenched hadron 
spectrum: one using the Wilson
$\kappa_{cW}$ definition, and the other using a definition which arises
by requiring that parity mixing, induced by the twisted mass term,
vanish in a physical basis \cite{dynamical2}.

Our spectrum calculation includes pseudoscalar mesons, vector mesons,
and baryons with spin-parity $(1/2)^\pm$ and $(3/2)^\pm$ for a 
degenerate (u,d) quark doublet. Computations
were carried out at two lattice spacings and four quark masses.

Flavor breaking induced by the chiral-flavor rotation in tmLQCD should be
suppressed by powers of the lattice spacing, but need not exactly vanish for
a nonzero lattice spacing.  Such an effect can be observed in principle
as a mass splitting between charged and neutral pions, though the computation
of a neutral pion mass involves diagrams with disconnected quark lines and
is therefore expensive. Flavor breaking can also arise in a baryon multiplet,
such as the isospin-3/2 $\Delta(1232)$, which contains states that are not
related
by u and d quark interchange. In this work, we calculate this mass splitting 
which does not involve any disconnected quark line contribution.

Section \ref{sec:action} contains a brief overview of tmLQCD and the
determination of the maximal twist angle is discussed. Details of the 
simulations and the results are given in Sec. \ref{sec:simul}. The discussion
of flavor breaking is given separately is Sec. \ref{sec:flavor}, and
Sec. \ref{sec:outlook} contains a summary.

\section{The action at maximal twist}\label{sec:action}

The Lagrangian of continuum QCD for two massless quark flavors is invariant
under the transformation
\begin{equation}\label{fieldtwist}
\psi(x) \to e^{i\omega\gamma_5\tau^3/2}\psi(x),
\end{equation}
where $\tau^3$ is the third Pauli matrix in flavor space and $\omega$
is called the twist angle.
When a degenerate quark mass $m_q$ is added, the fermionic part of the
Lagrangian is no longer
invariant.  After the transformation in Eq.~(\ref{fieldtwist}) it becomes
(in Euclidean notation)
\begin{equation}\label{continuumL}
{\cal L}_F(x) =
      \bar\psi(x)\left(D\!\!\!\!/+m+i\mu\gamma_5\tau^3\right)\psi(x),
\end{equation}
where $D_\nu$ is the covariant derivative and the parameters $m$ and $\mu$
will be called, respectively, the standard
mass parameter and the twisted mass parameter.  They are given by
\begin{eqnarray}
m &=& m_q\cos\omega, \\
\mu &=& m_q\sin\omega.
\end{eqnarray}
Two choices for the twist angle $\omega$ are of particular
interest: $\omega=0$ causes $\mu$ to vanish and gives the QCD Lagrangian in
its usual form, while $\omega=\pi/2$ causes $m$ to vanish and is referred to 
as maximal twist. Note that for the continuum action the twist can not effect 
physics since it can be removed by a field redefinition.

One immediately sees that vector and axial currents get mixed by a general
rotation.  For example the charged currents,
\begin{eqnarray}
V_\nu^-(x) &\equiv& \bar u(x)\gamma_\nu d(x), \\
A_\nu^-(x) &\equiv& \bar u(x)\gamma_\nu\gamma_5 d(x),
\end{eqnarray}
transform as follows:
\begin{eqnarray}
V_\nu^-(x) &\to& \tilde V_\nu^-(x)\cos\omega - i\tilde A_\nu^-(x)\sin\omega,
 \label{defV} \\
A_\nu^-(x) &\to& \tilde A_\nu^-(x)\cos\omega - i\tilde V_\nu^-(x)\sin\omega.
\end{eqnarray}
A tilde has been added to all currents constructed from fields in the twisted basis.
Notice that the vector and axial currents don't mix at maximal twist, but
their roles get interchanged.

TmLQCD is obtained when the transformation of
Eq.~(\ref{fieldtwist}) is applied to the mass term of the Wilson lattice QCD
action \cite{tmLQCD}.
The fermion action used in the present work is in the twisted basis,
\begin{equation}\label{action}
S_F[\psi,\bar\psi,U] = a^4\sum_x\bar\psi(x)\left(M+i\mu\gamma_5\tau^3
     +\frac{1}{2}\sum_\nu\left(\gamma_\nu\nabla_\nu+\gamma_\nu\nabla_\nu^*
     -a\nabla_\nu^*\nabla_\nu\right)\right)\psi(x),
\end{equation}
where
\begin{eqnarray}
\nabla_\nu\psi(x) &=& U_\nu(x)\psi(x+a\hat\nu) - \psi(x), \\
\nabla_\nu^*\psi(x) &=& \psi(x) - U_\nu^\dagger(x-a\hat\nu)\psi(x-a\hat\nu).
\end{eqnarray}
In the lattice theory, both the mass term and the chiral
symmetry breaking Wilson term lack invariance under Eq.~(\ref{fieldtwist}) so,
unlike the continuum case of Eq.~(\ref{continuumL}), the chiral twist can not 
be removed from the action by a field redefinition. Essentially tmLQCD
generalizes the Wilson fermion action by introducing a relative chiral 
phase between the mass term and the Wilson term. 

The mass parameter $M$ in Eq.~(\ref{action}) undergoes both additive and 
multiplicative renormalization, but the twisted mass parameter $\mu$ is only
renormalized multiplicatively. It is usual \cite{tmLQCD,FreRos}
to divide $M$ into a mass 
shift $M_{sh}$ and a quark mass excess $m$ which plays the same role as the 
standard mass parameter in Eq.~(\ref{continuumL}). In particular it is the 
quark mass excess which, along with $\mu$, determines the twist angle
\cite{tmLQCD,FreRos}. Our quark propagator calculation will be done using the
fermion action of Eq.~(\ref{action}) in the twisted basis so we dispense with
this
separation into mass shift and quark mass excess and simply quote $M$ as the 
standard mass parameter.

Various definitions of maximal twist are possible, and they will generally
differ by lattice spacing effects.
The Wilson $\kappa_{cW}$ definition of maximal twist \cite{FreRos} is 
obtained from the standard (unimproved) Wilson
action ($\mu=0$) by computing the pseudoscalar meson mass as a function of
$M$ (equivalently the hopping parameter $\kappa$), 
then extrapolating to vanishing meson mass.  This critical standard
mass parameter $M_c$ is used for the mass shift $M_{sh}$ at all $\mu$
and simulations at $M = M_c$, $\mu\neq0$ correspond to maximal twist
in this definition. In terms of the hopping parameter,
the critical standard mass parameter is
\begin{equation}\label{kappacWdef}
aM_c = \frac{1}{2\kappa_{cW}} - 4.
\end{equation}
This Wilson $\kappa_{cW}$ definition of maximal twist has been used previously 
in simulations\cite{Jansen,GoingChiral,pionff}, but studies of $O(a)$ 
improvement in twisted mass chiral perturbation theory recommend against
it \cite{AokiBar,SharpeWu2}.  Simulations with this definition are discussed in
the present work as well, and are compared to a definition of maximal twist
arising from considerations of parity conservation.

The tmLQCD action expressed in terms of twisted fields has a parity violating
mass term. This term may be removed by a field redefinition and then the 
parity violation is associated with the Wilson term. The resulting form of the
action is said to be in the physical basis \cite{FreRos}.
The parity conservation definition of the twist angle comes from demanding 
the absence of a mixed pseudoscalar-vector matrix element in the physical basis
\cite{dynamical2,SharpeWu2},
\begin{equation}
\sum_{\vec x}\left<V_\nu^-(\vec x,t)P^+(0)\right> = 0,
\end{equation}
with
\begin{equation}
P^+(x) \equiv \bar d(x)\gamma_5u(x).
\end{equation}
Transforming this according to Eq.~(\ref{defV}) and noting that
$P^+(x)$ is invariant under the transformation of Eq.~(\ref{fieldtwist}),
we arrive at
\begin{equation}\label{twistangle}
\tan\omega = \frac{i\sum_{\vec x}\left<\tilde V_\nu^-(\vec x,t)P^+(0)\right>}
                  {\sum_{\vec x}\left<\tilde A_\nu^-(\vec x,t)P^+(0)\right>}.
\end{equation}
where a tilde again denotes currents constructed from fields in the twisted basis.
This definition is essentially independent of Euclidean time $t$, as will be
seen explicitly below.
In this work, we compute the hadron spectrum for four values of $\mu$,
with $M$ tuned in each case such that $\omega$ determined by
Eq.~(\ref{twistangle}) is $\pi/2$.

\section{Simulations}\label{sec:simul}
\subsection{Maximal twist from Wilson $\kappa_{cW}$}\label{sec:Wilsontwist}

In this section, the quenched hadron spectrum is obtained by using the Wilson
$\kappa_{cW}$ definition of maximal twist.
A standard heatbath algorithm with the Wilson plaquette action
was used to create two quenched gauge field
ensembles of 1000 configurations each, one at $\beta=5.85$
and the other at $\beta=6.0$, corresponding to lattice spacings near
$a=0.123$ fm and $a=0.1$ fm respectively \cite{Jansen}.

Twisted mass quark propagators are computed from the action in
Eq.~(\ref{action})
with a periodic boundary condition in all directions
using a 1-norm quasi-minimal residual algorithm. 
The standard mass parameter for each $\beta$ is set to the value
obtained using $\kappa_{cW}$ from Ref.~\cite{Jansen}, 
and four different values for the twisted mass parameter
are chosen which span the range of quark masses between $m_s$ and $m_s/6$.
($m_s$ denotes the strange quark mass.) From the point of view of 
numerical convergence of the propagator calculation, it would be possible
to push the calculation to smaller quark masses. However, that would 
probably require even higher statistics to get meaningful results so
for our first pass we have not tried to probe for the ultimate small 
quark mass limit. As it is, we achieve a pseudoscalar-meson to vector-meson 
mass ratio of about $1/3$. Numerical values of the
simulation parameters are displayed in Table~\ref{tab:parameters}.

To uncover any finite volume corrections, simulations were run at $\beta=5.85$
on both $16^3\times40$ and $20^3\times40$ lattices.  Differences in the
mass spectrum were found to be negligible, so only results from the larger
volume will be presented.  Since the smaller volume was near (2 fm)$^3$,
we chose a comparable physical volume for the $\beta=6.0$ simulations.

The ground state masses of the pion, $\rho$ meson, $(1/2)^\pm$ baryons
and $(3/2)^\pm$ baryons are obtained from the following operators,
defined in the physical quark basis:
\begin{eqnarray}
\pi^\pm &=& \bar\psi\gamma_5\tau^\mp\psi, \label{pionop} \\
\rho_i^\pm &=& \bar\psi\gamma_i\tau^\mp\psi, \label{rhoop} \\
\tilde\rho_i^\pm &=& \bar\psi\sigma_{i4}\tau^\mp\psi, \label{rhotildeop} \\
p &=& \epsilon_{abc}[u_a^TC\gamma_5d_b]u_c, \label{protonop} \\
n &=& \epsilon_{abc}[d_a^TC\gamma_5u_b]d_c, \\
\Delta_i^{++} &=& \epsilon_{abc}[u_a^TC\gamma_iu_b]u_c, \\
\Delta_i^+ &=& \frac{2}{\sqrt{3}}\epsilon_{abc}[u_a^TC\gamma_id_b]u_c
             + \frac{1}{\sqrt{3}}\epsilon_{abc}[u_a^TC\gamma_iu_b]d_c, \\
\Delta_i^0 &=& \frac{2}{\sqrt{3}}\epsilon_{abc}[d_a^TC\gamma_iu_b]d_c
             + \frac{1}{\sqrt{3}}\epsilon_{abc}[d_a^TC\gamma_id_b]u_c, \\
\Delta_i^- &=& \epsilon_{abc}[d_a^TC\gamma_id_b]d_c, \label{Deltaminusop}
\end{eqnarray}
where $C$ is the charge conjugation matrix, $i$ is a spatial Lorentz index,
and sums over color indices $a$, $b$ and $c$ are implied.
Since the $\rho_i$ operator tends to produce a somewhat noisy correlator in
tmLQCD simulations, the operator $\tilde\rho_i$ is also considered.

As is well known, the $\Delta_i$ operators contain a spin 1/2 admixture.
At zero momentum, the correlator for spatial Lorentz indices is\cite{BenDM}
\begin{equation}
\sum_{\vec x}\left<0\left|T\left(\Delta_j(\vec x,t)\bar\Delta_i(\vec 0,0)
\right)\right|0\right> \equiv G_{ji}(t)
= \left(\delta_{ji}-\frac{1}{3}\gamma_j\gamma_i\right)G_{3/2}(t)
 + \frac{1}{3}\gamma_j\gamma_iG_{1/2}(t).
\end{equation}
The spins are easily separated,
\begin{eqnarray}
G_{1/2}(t) &=& \frac{1}{3}{\rm Tr}[G_{ji}(t)\gamma_i\gamma_j], \\
G_{3/2}(t) &=& \frac{1}{6}{\rm Tr}[G_{ji}(t)\gamma_j\gamma_i+C_{ii}(t)],
\end{eqnarray}
where repeated indices are summed.
We will use this $G_{3/2}(t)$ to determine the masses of spin 3/2 baryons.

To get twisted versions of the operators in
Eqs.~(\ref{pionop}-\ref{Deltaminusop}), one simply
applies the twist from Eq.~(\ref{fieldtwist}).
In practice the simulation is done by computing the tmLQCD quark propagators 
in the twisted basis using Eq.~(\ref{action}) and then applying 
Eq.~(\ref{fieldtwist}) to both ends. The correlators then involve the 
operators unchanged in the physical basis.

As discussed in Ref.~\cite{SBO}, baryon two-point correlators couple to both
parities even without twisting.  For periodic boundary conditions, the
correlator is
\begin{equation}
G(t) = \sum_{s=\pm}
\left[(1+s\gamma_4)e^{-M_st}+(1-s\gamma_4)e^{-M_s(T-t)}\right]sG^{(s)}(t),
\end{equation}
where $T$ is the temporal extent of the lattice, and
$s$ sums over positive and negative parity.
This expression is directly applicable to our tmLQCD simulations at maximal
twist as well.
We can therefore discuss the ground state masses of both parities
from the baryon operators in Eqs.~(\ref{protonop}-\ref{Deltaminusop}).

Throughout this work, all hadron masses are obtained from three-exponential
fits to the hadron correlators using all time slices except the source.  
Statistical uncertainties are
obtained from the bootstrap method with replacement, where the number of
bootstrap ensembles is three times the number of data points in the original
ensemble.

\begin{figure}[tb]
\vspace{4mm}
\includegraphics[width=10cm]{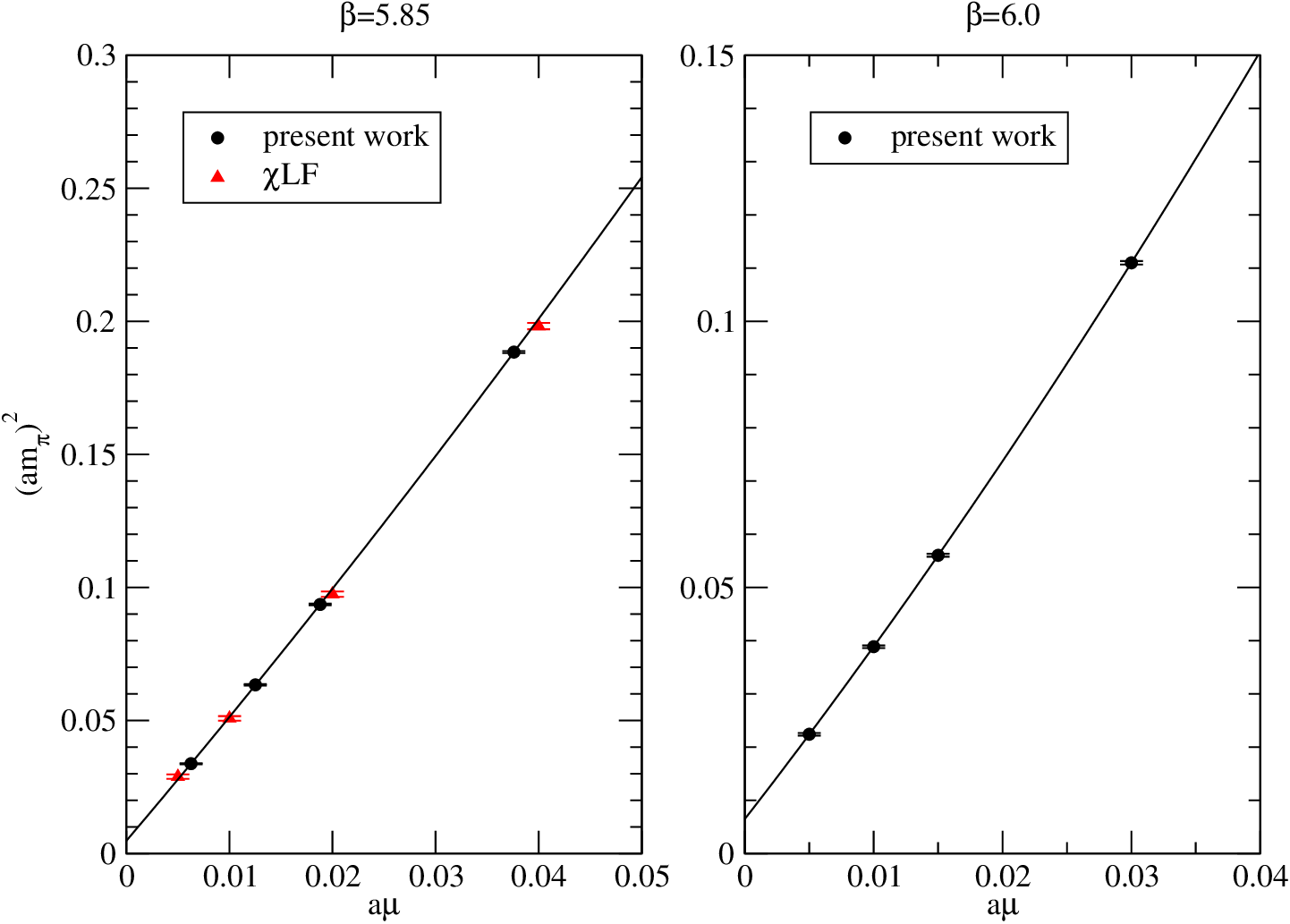}
\caption{Pseudoscalar meson mass squared as a function of the twisted
         mass parameter.  The standard mass parameter was held fixed to the 
         value obtained from the
         Wilson $\kappa_{cW}$ definition of maximal twist, 
         Eq.~(\protect\ref{kappacWdef}). Data labelled by
         ``$\chi$LF'' are taken from Ref.~\protect\cite{GoingChiral}.}
\label{fig:mpivsmuwilson}
\end{figure}

Our pion mass squared is obtained from the configuration average of
$\pi^+$ and $\pi^-$
and is plotted versus $\mu$ in Fig.~\ref{fig:mpivsmuwilson}.
Excellent agreement is obtained with the $\beta=5.85$ tmLQCD results from 
Ref.~\cite{GoingChiral}.
A least squares fit of our data to the form
\begin{equation}\label{extrapfunc}
(am_\pi)^2 = A + B(a\mu) + C(a\mu)^2
\end{equation}
is also displayed.  Notice that the fitted curve does not lead to a vanishing
pion mass for $\mu=0$, an effect also noted in Ref.~\cite{GoingChiral}.
There could be various reasons for this but one's
first guess might be that it is due to lattice spacing errors in the
determination of $\kappa_{cW}$ and of the twist angle.  We will revisit 
this issue, and also consider the pion decay constant, in
Sec.~\ref{sec:paritytwist} in the context of an alternate
definition of maximal twist.

\begin{figure}[tb]
\vspace{4mm}
\includegraphics[width=10cm]{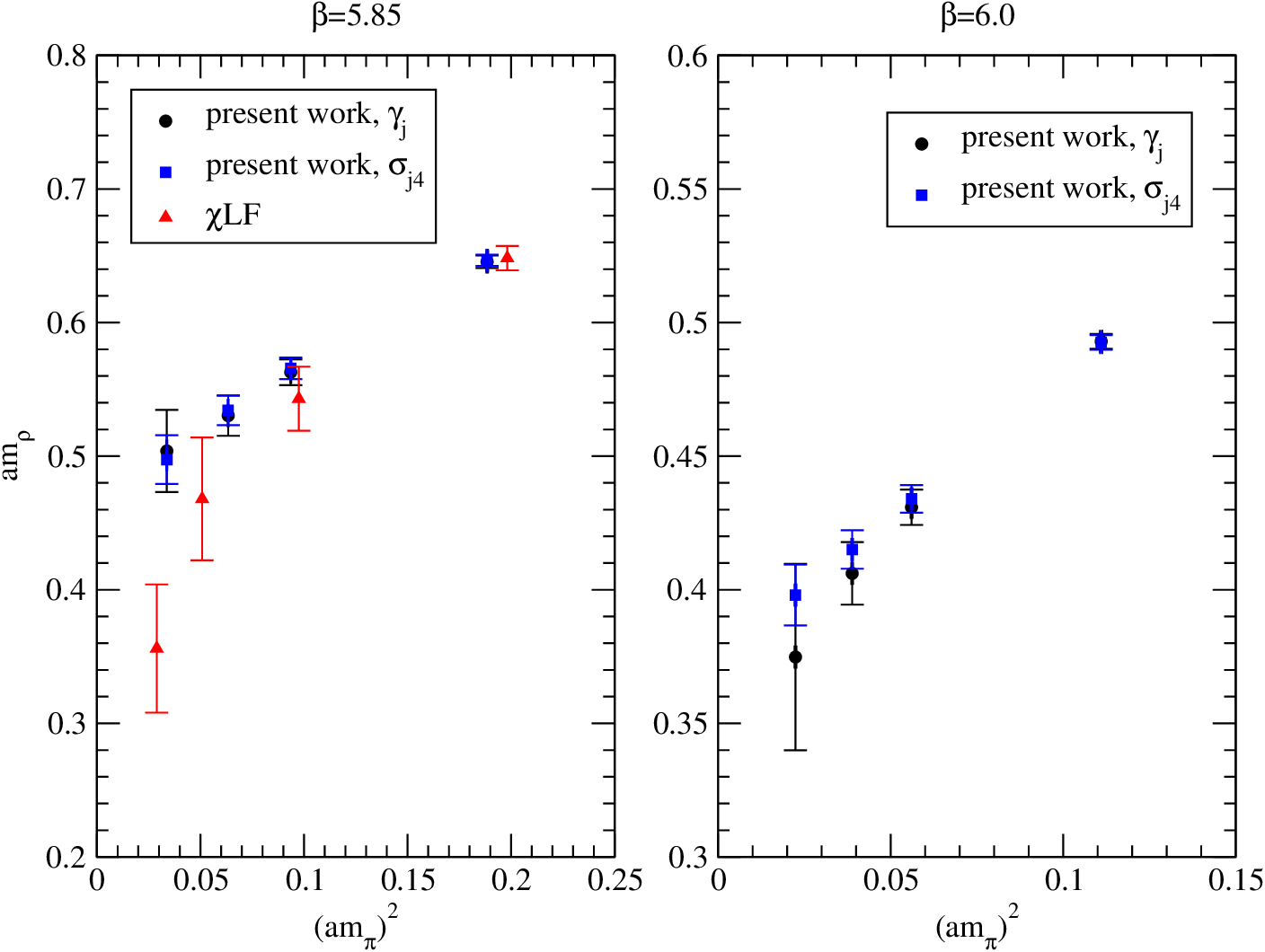}
\caption{Vector meson mass as a function of the pseudoscalar meson mass
         squared.
         The standard mass parameter was held fixed to the value obtained from the
         Wilson $\kappa_{cW}$ definition of maximal twist, 
         Eq.~(\protect\ref{kappacWdef}). Data labelled by
         ``$\chi$LF'' are taken from Ref.~\protect\cite{GoingChiral}.}
\label{fig:mrhovsmpiwilson}
\end{figure}

Fig.~\ref{fig:mrhovsmpiwilson} is a plot of vector meson mass, averaged over
$\rho^+$ and $\rho^-$ and over spatial Lorentz indices, versus pseudoscalar
mass squared,
again with tmLQCD results from Ref.~\cite{GoingChiral} for comparison.
It is interesting to see that the operator from Eq.~(\ref{rhotildeop}) has
consistently smaller uncertainties than the traditional vector operator of
Eq.~(\ref{rhoop}).  It is also noted that a linear extrapolation to
vanishing pion mass squared produces a vector mass that agrees with the
experimental $\rho$ meson mass, at both $\beta$ values.

\begin{figure}[tb]
\vspace{4mm}
\includegraphics[width=10cm]{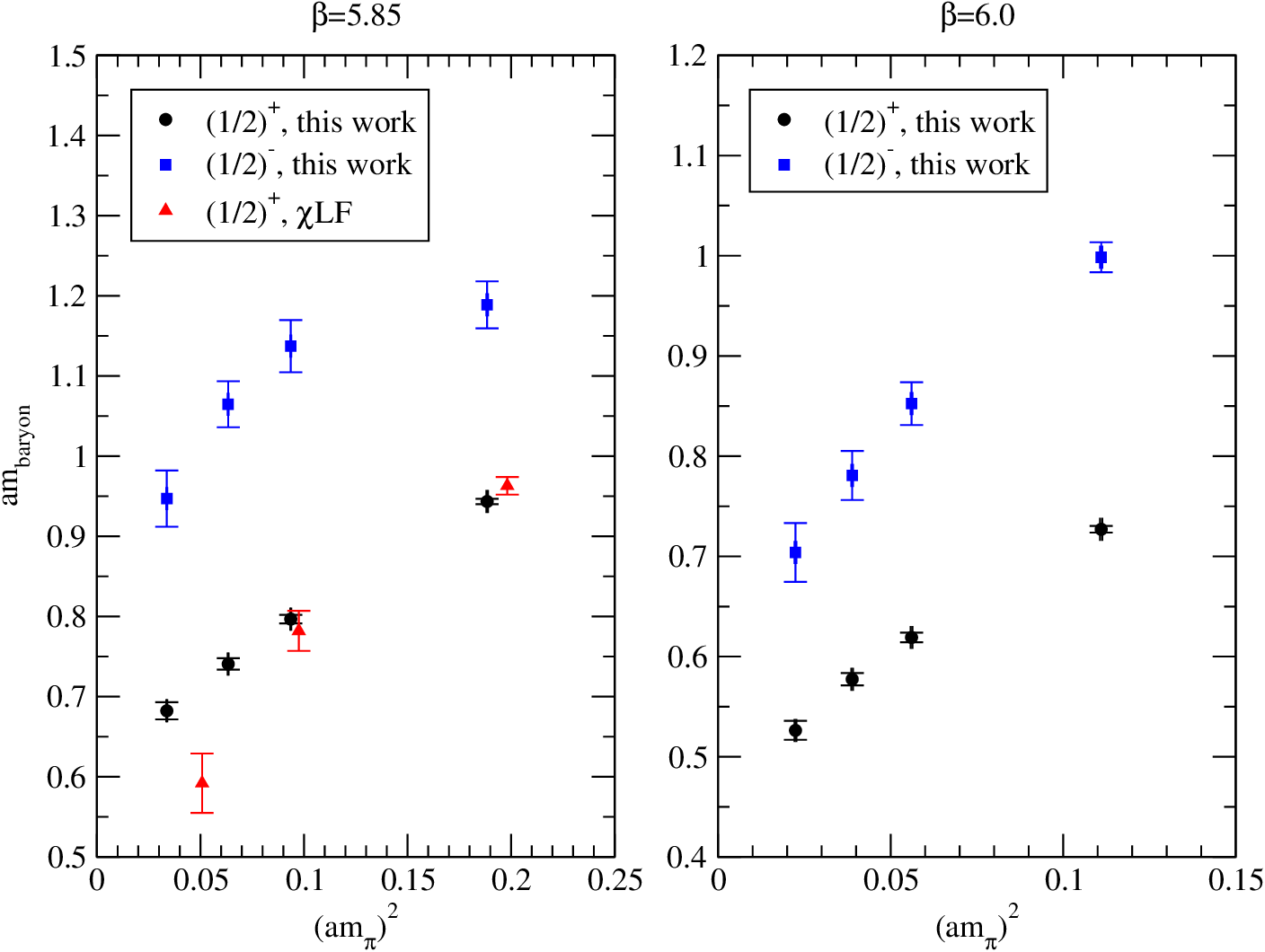}
\caption{Spin 1/2 baryon masses (both parities) as functions
         of the pseudoscalar meson mass squared.
         The standard mass parameter was held fixed to the value obtained from the
         Wilson $\kappa_{cW}$ definition of maximal twist, 
         Eq.~(\protect\ref{kappacWdef}).
         Data labelled by
         ``$\chi$LF'' are taken from Ref.~\protect\cite{GoingChiral}.}
\label{fig:mNvsmpiwilson}
\end{figure}

\begin{figure}[tb]
\vspace{4mm}
\includegraphics[width=10cm]{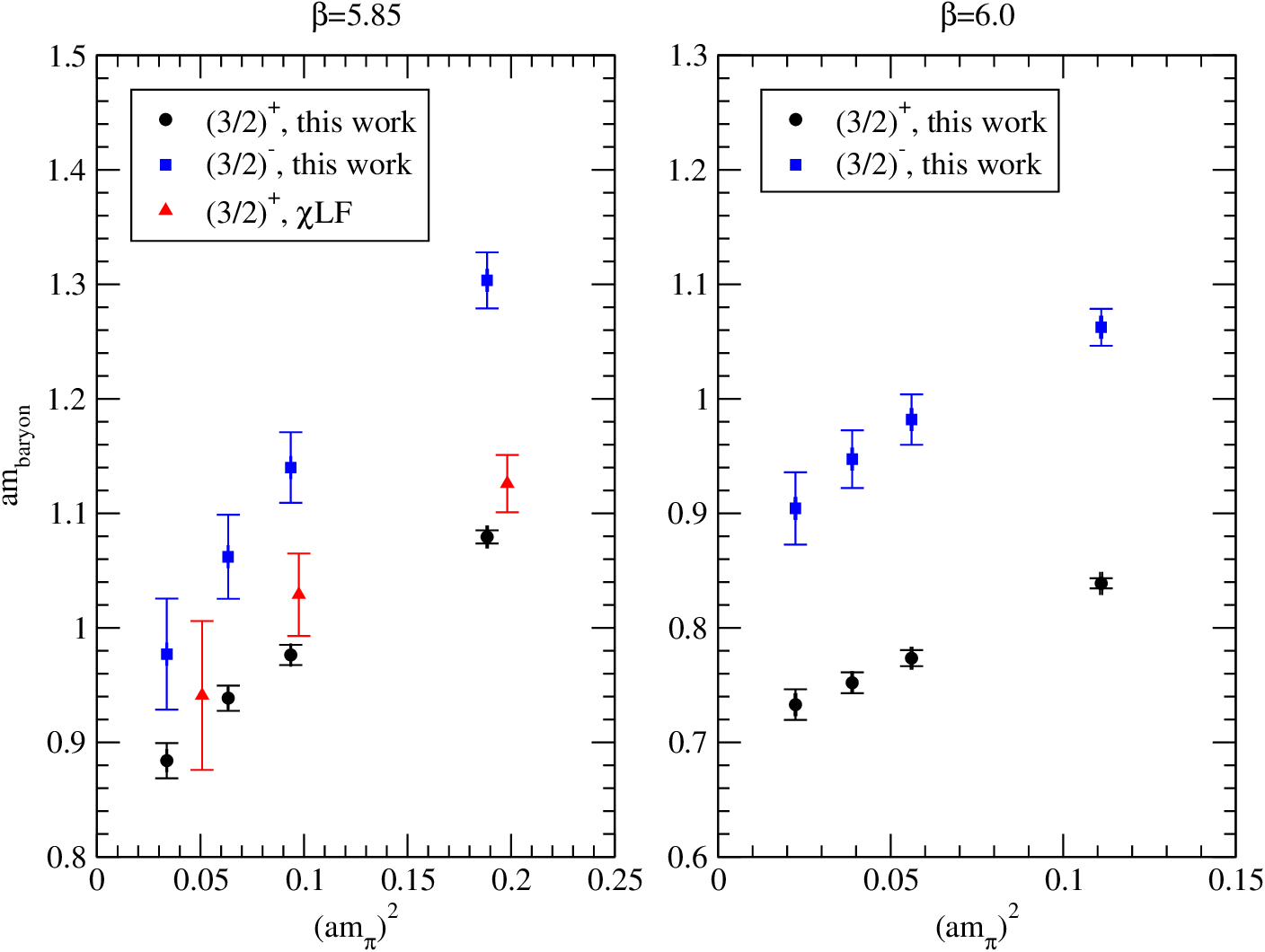}
\caption{Spin 3/2 baryon masses (both parities) as functions
         of the pseudoscalar meson mass squared.
         These data are averages of the $\Delta^{++}$ and $\Delta^-$ masses.
         The standard mass parameter was held fixed to the value obtained from the
         Wilson $\kappa_{cW}$ definition of maximal twist, 
         Eq.~(\protect\ref{kappacWdef}).
         Data labelled by
         ``$\chi$LF'' are taken from Ref.~\protect\cite{GoingChiral}.}
\label{fig:mDvsmpiwilson}
\end{figure}

Figures~\ref{fig:mNvsmpiwilson} and \ref{fig:mDvsmpiwilson}
show the baryon masses obtained from our fits along with positive parity results
from Ref.~\cite{GoingChiral} for comparison.
In these plots, our nucleon is the configuration average of proton and
neutron, while our $\Delta$ is the configuration average of
$\sum_{i=1}^3\Delta^{++}_i$ and $\sum_{i=1}^3\Delta^-_i$.
Extrapolations of the positive parity baryons are within about 10\% of the
experimental masses. The connection of negative parity data to experiment
is less clear.

\begin{figure}[tb]
\vspace{4mm}
\includegraphics[width=10cm]{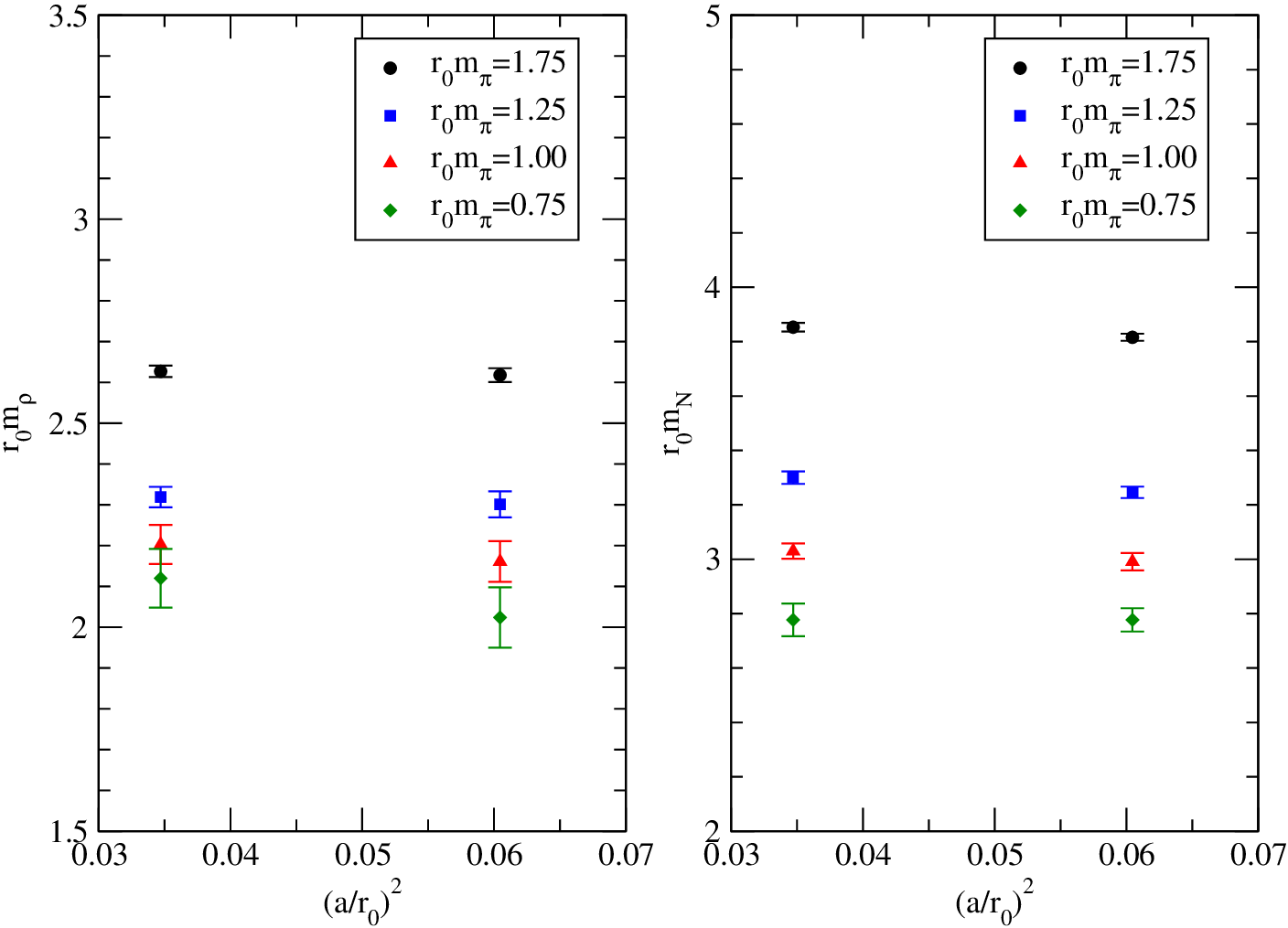}
\caption{Vector meson and $(1/2)^+$ baryon masses as functions
         of lattice spacing squared.  Axes are in units of $r_0=0.5$ fm.
         The calculation was done using the
         Wilson $\kappa_{cW}$ definition of maximal twist, 
         Eq.~(\protect\ref{kappacWdef}).}
\label{fig:scalingwilson}
\end{figure}

The $O(a)$ improvement of tmLQCD should result in better scaling between
lattices at $\beta=5.85$ and $\beta=6.0$ than would be obtained from the
Wilson action.  This was studied for the mass spectrum in Ref.~\cite{Jansen}
and discussed for the pion form factor in Ref.~\cite{pionff}.  Our present
data for the $\rho$ and nucleon masses have only a very small dependence on
lattice spacing relative to the statistical uncertainties, as displayed in
Figure~\ref{fig:scalingwilson}.

\subsection{Maximal twist from parity conservation}\label{sec:paritytwist}

\begin{figure}[tb]
\vspace{4mm}
\includegraphics[width=10cm]{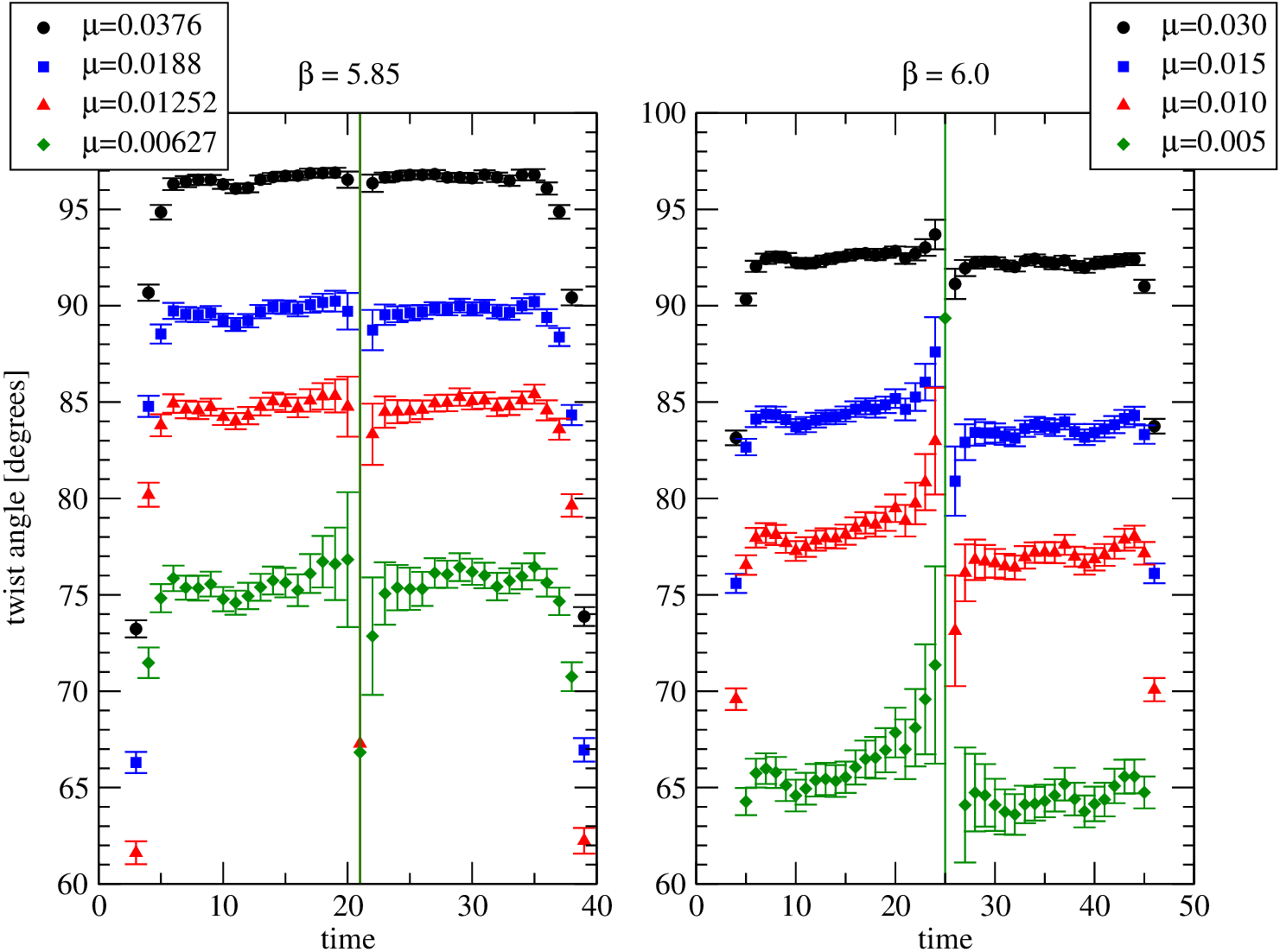}
\caption{The twist angle obtained from Eq.~(\protect\ref{twistangle})
           at different lattice time slices while using the
         Wilson $\kappa_{cW}$ definition of maximal twist, 
         Eq.~(\protect\ref{kappacWdef}).}
\label{fig:anglewilson}
\end{figure}

The Wilson definition of maximal twist used in Sec.~\ref{sec:Wilsontwist}
relies on Eq.~(\ref{kappacWdef}) with the numerical values of $\kappa_{cW}$
taken from Ref.~\cite{Jansen}.  In this section, maximal twist will be
defined by tuning using Eq.~(\ref{twistangle}).  To show that the two options 
are numerically quite different, Fig.~\ref{fig:anglewilson} displays
$\omega$ as obtained from Eq.~(\ref{twistangle}) when the standard and
twisted mass parameters keep the values used in Sec.~\ref{sec:Wilsontwist}.
The source is at (Euclidean) time 1 in this plot.
Each $\mu$ value is found to correspond to a different twist angle $\omega$,
and the angles range from 65$^\circ$ to more than 90$^\circ$.

In light of cautionary notes against relying on the Wilson $\kappa_{cW}$
definition of maximal twist\cite{AokiBar,SharpeWu2}, we will now tune 
$\omega$ to $\pi/2$ according to Eq.~(\ref{twistangle}) and recompute the 
mass spectrum. 
Parameters used in the simulations are listed in Table~\ref{tab:parameters}.
Notice in particular that these results use 300 gauge field configurations 
as opposed to the 1000 configurations used for the $\kappa_{cW}$ definition 
of maximal twist in Sec.~\ref{sec:Wilsontwist}.

\begin{figure}[tb]
\vspace{4mm}
\includegraphics[width=10cm]{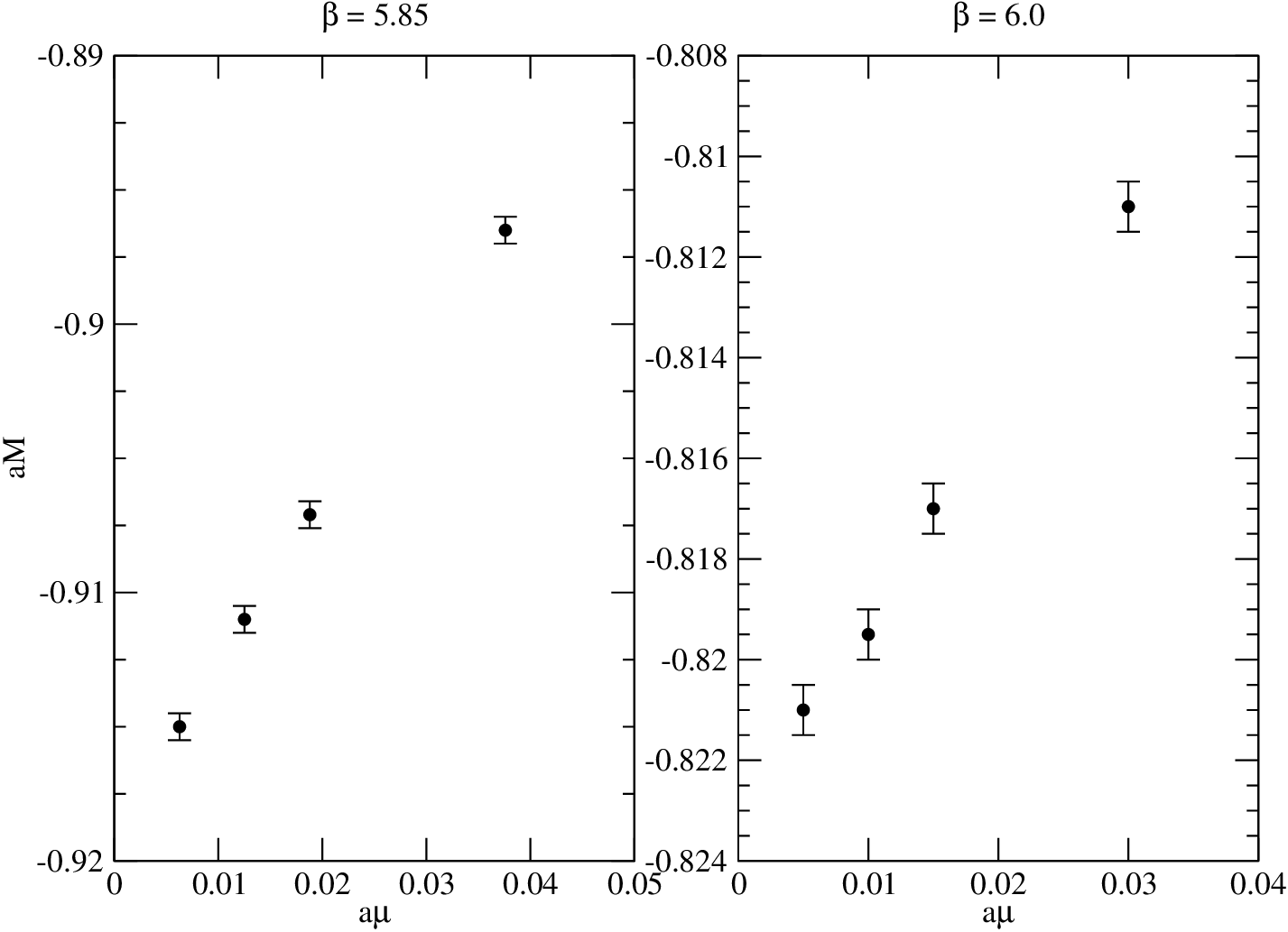}
\caption{The standard mass parameter as a function of the twisted mass
         parameter, tuned such that the twist angle $\omega$ of
         Eq.~(\protect\ref{twistangle}) is $\pi/2$.}
\label{fig:angleparity}
\end{figure}

Fig.~\ref{fig:angleparity} is a plot of the standard mass parameter $M$ tuned
such that $\omega=\pi/2$ for each twisted mass parameter.
The dependence of $aM$ on $a\mu$ is seen to be quite linear with a
distinctly nonzero slope.

\begin{figure}[tb]
\vspace{4mm}
\includegraphics[width=10cm]{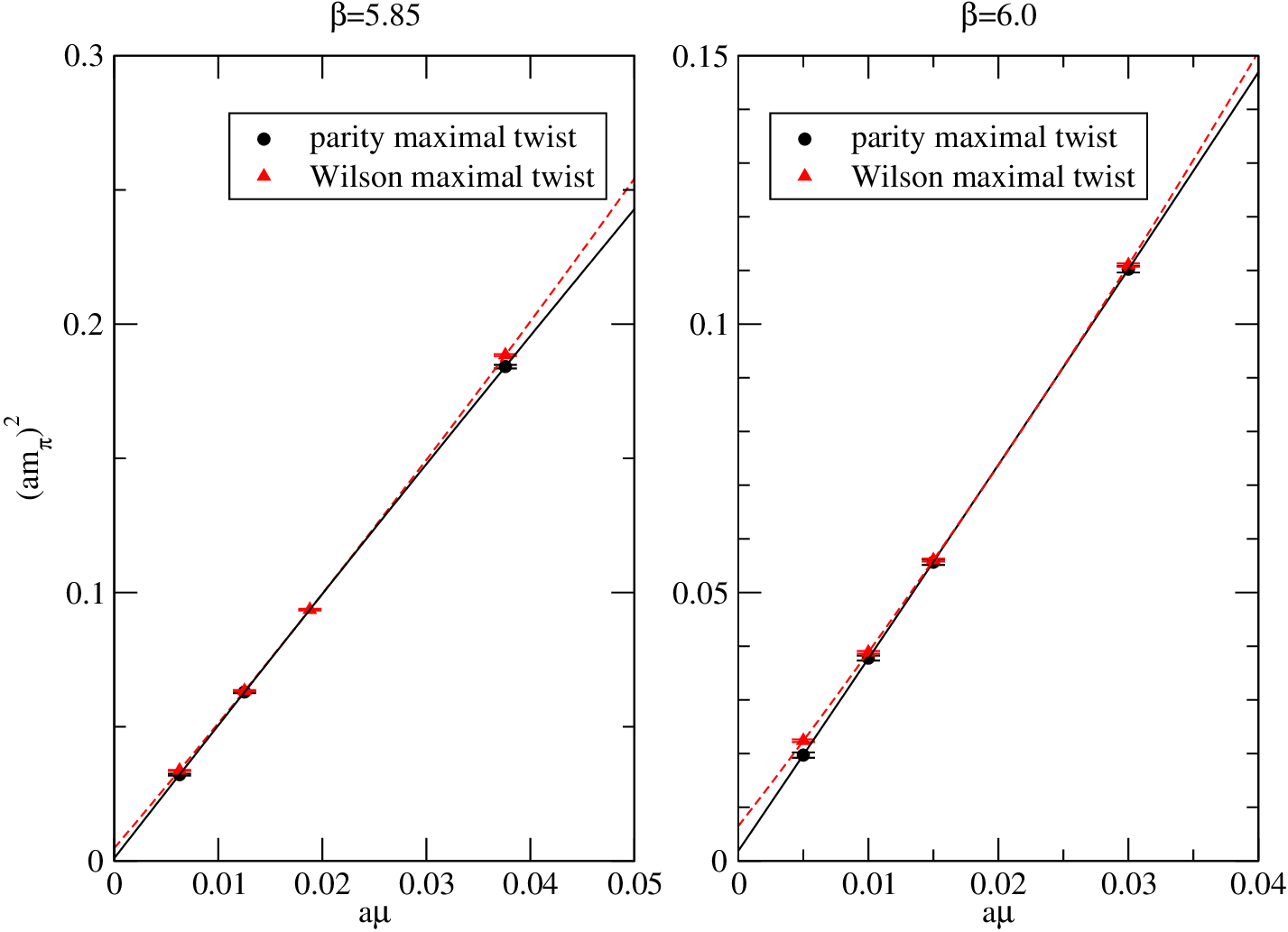}
\caption{Pseudoscalar meson mass squared as a function of the twisted
         mass parameter, calculated using the parity conservation definition of maximal
         twist: $\omega=\pi/2$ in Eq.~(\protect\ref{twistangle}).
         Data calculated using the Wilson $\kappa_{cW}$ definition, Fig.~\protect\ref{fig:mpivsmuwilson},
         are shown for comparison.}
\label{fig:mpivsmuparity}
\end{figure}

The pseudoscalar meson mass squared is plotted for both definitions of
maximal twist in Fig.~\ref{fig:mpivsmuparity}.  The least squares fit of each
to a quadratic is also shown.  It is evident that the parity conservation 
definition of maximal twist, i.e. Eq.~(\ref{twistangle}), produces a 
smaller pion mass for sufficiently small $\mu$, thus demonstrating that 
the $\kappa_{cW}$ definition did not precisely correspond to $\omega=\pi/2$ 
for this observable.  In addition, Fig.~\ref{fig:mpivsmuparity} indicates 
that the parity conservation definition of maximal twist leads to 
less curvature in the $m_\pi^2$  versus $\mu$ extrapolation and a 
smaller intercept for $m_\pi^2$ at $\mu=0$.

The curvature of $m_\pi^2$ as a function of quark mass and the non-zero
value at zero quark mass are features commonly found in lattice simulations.
Higher order lattice spacing errors may lead to a residual pion mass even
after tuning to maximal twist to remove  $O(a)$ effects. Finite volume effects
can not be discounted, although our simulation at $\beta=5.85$ in 
Sec. \ref{sec:Wilsontwist} did not give any indication of such a problem.
In addition, the simple extrapolation function Eq.~(\ref{extrapfunc}), although
consistent with simulation data, may not be appropriate at very small 
quark masses. Quenched chiral perturbation theory suggests a non-analytic
behavior of $m_\pi^2$ at zero quark mass \cite{BerGol, Sharpe92, MSS}.

The chiral fits of Chen {\it et al.}~\cite{chen} to a large set of pion 
masses calculated at very small quark masses (using the overlap 
fermion approach) are very instructive. They show how quenched chiral
logarithms can conspire to fit simulation data with an essentially 
linear quark mass behavior in $m_\pi^2$ with the pion mass dipping down 
to zero only at extremely small quark masses. That paper also contains
many references to earlier searches for quenched chiral logarithms.

It may be interesting to look for quenched chiral logarithms
with tmLQCD. However Ref.~\cite{chen} suggests much more data at
quark masses smaller than those considered in this work would be needed.
If one has to tune the standard mass parameter as a function of the twisted
mass $\mu$, this would add to the overall cost of such a calculation.

\begin{figure}[tb]
\vspace{4mm}
\includegraphics[width=10cm]{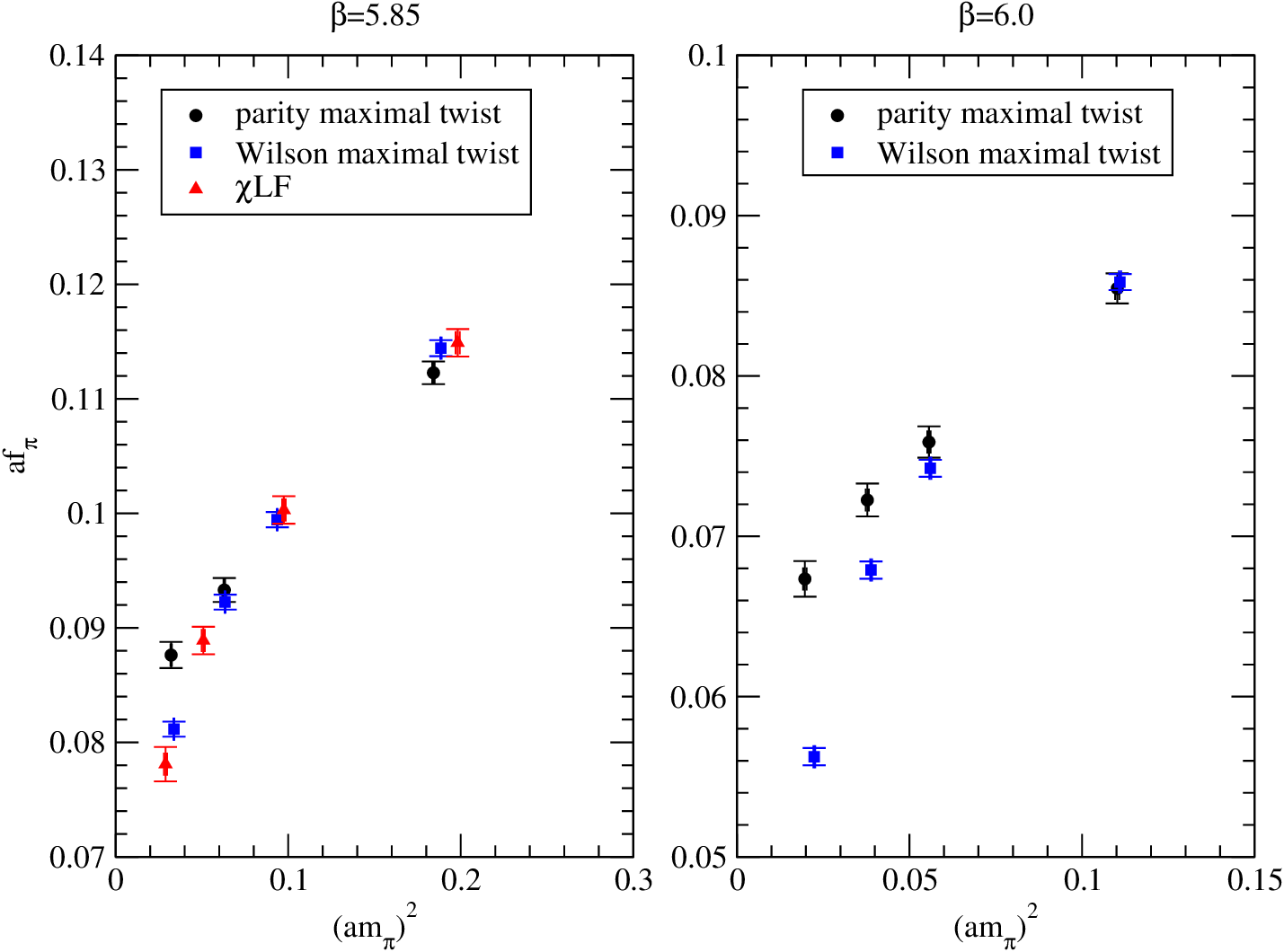}
\caption{Pseudoscalar meson decay constant as a function of the pseudoscalar
         meson mass
         squared, calculated using both the parity conservation definition and
         the Wilson $\kappa_{cW}$ definition of maximal twist.
         Data labelled by
         ``$\chi$LF'' are taken from Ref.~\protect\cite{GoingChiral}.}
\label{fig:decayconstant}
\end{figure}

\begin{figure}[tb]
\vspace{4mm}
\includegraphics[width=10cm]{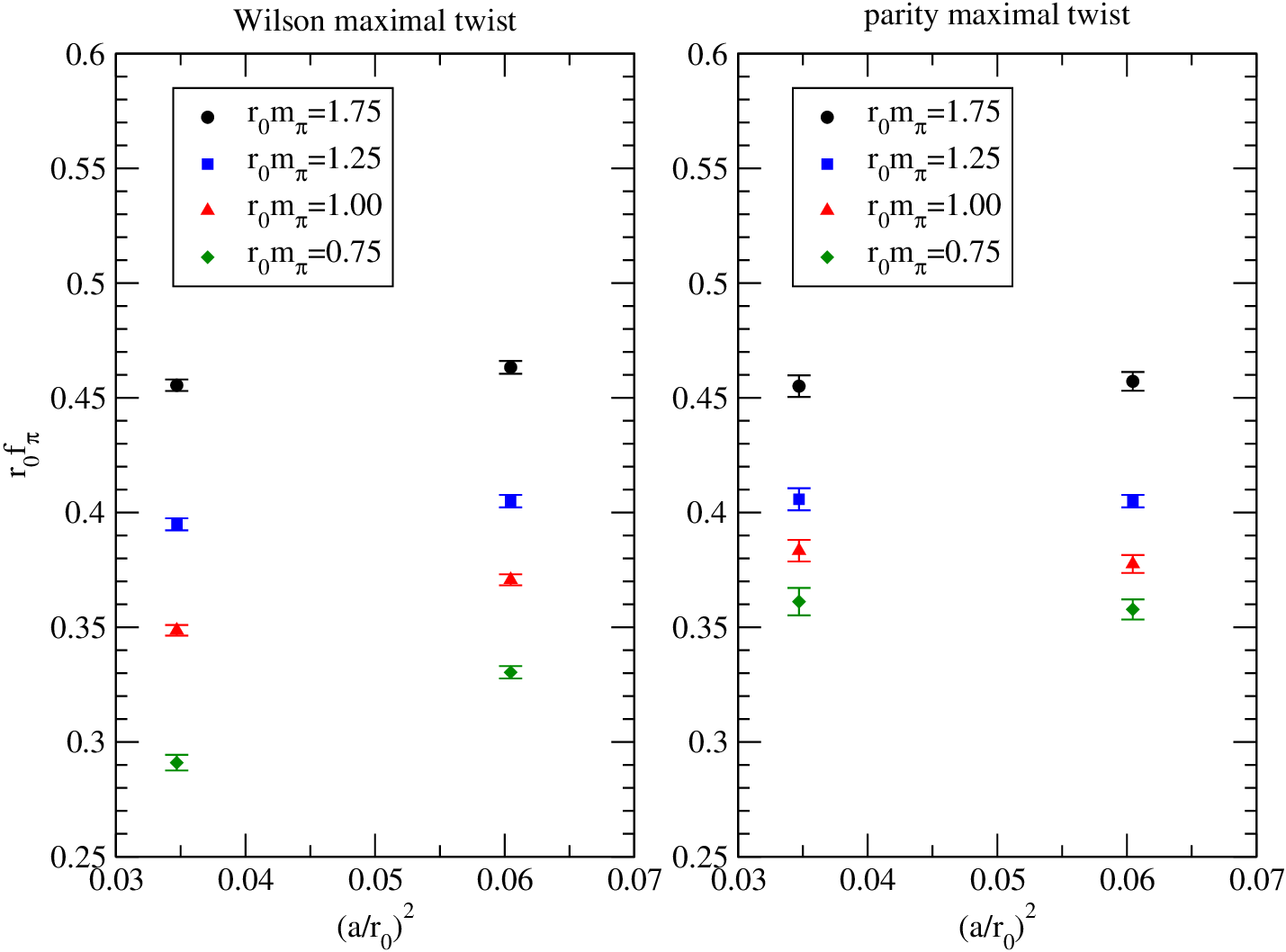}
\caption{Pseudoscalar meson decay constant as a function
         of lattice spacing squared, calculated using both the
         parity conservation definition and
         the Wilson $\kappa_{cW}$ definition of maximal twist.
         Axes are in units of $r_0=0.5$ fm.}
\label{fig:fpiscaling}
\end{figure}

In addition to the pion mass, our simulations provide a determination of the
pion decay constant.  The ``indirect'' method of Ref.~\cite{GoingChiral} is
particularly convenient,
\begin{equation}
f_\pi = \frac{2\mu}{m_\pi^2}\bigg|\left<0|P^\pm|\pi\right>\bigg|,
\end{equation}
where the normalization is such that the physical value is
$f_\pi\approx130$ MeV.
Fig.~\ref{fig:decayconstant} compares the results for the two definitions of
maximal twist.  Numerical results from Ref.~\cite{GoingChiral} using the
Wilson $\kappa_{cW}$ definition are also displayed.
Much like the pion mass squared, the decay constant has significantly less
curvature in the quark mass when the parity conservation definition of maximal
twist is used, and is consistent
with a linear function of $m_\pi^2$ as expected from chiral
perturbation theory.
In contrast, data from the Wilson $\kappa_{cW}$ definition are clearly not
linear and do not approach a realistic value.
Fig.~\ref{fig:fpiscaling} shows that the decay constant is independent
of lattice spacing for all quark masses within uncertainties for the parity
definition, while the $\kappa_{cW}$ definition does not scale as well.

\begin{figure}[tb]
\vspace{4mm}
\includegraphics[width=10cm]{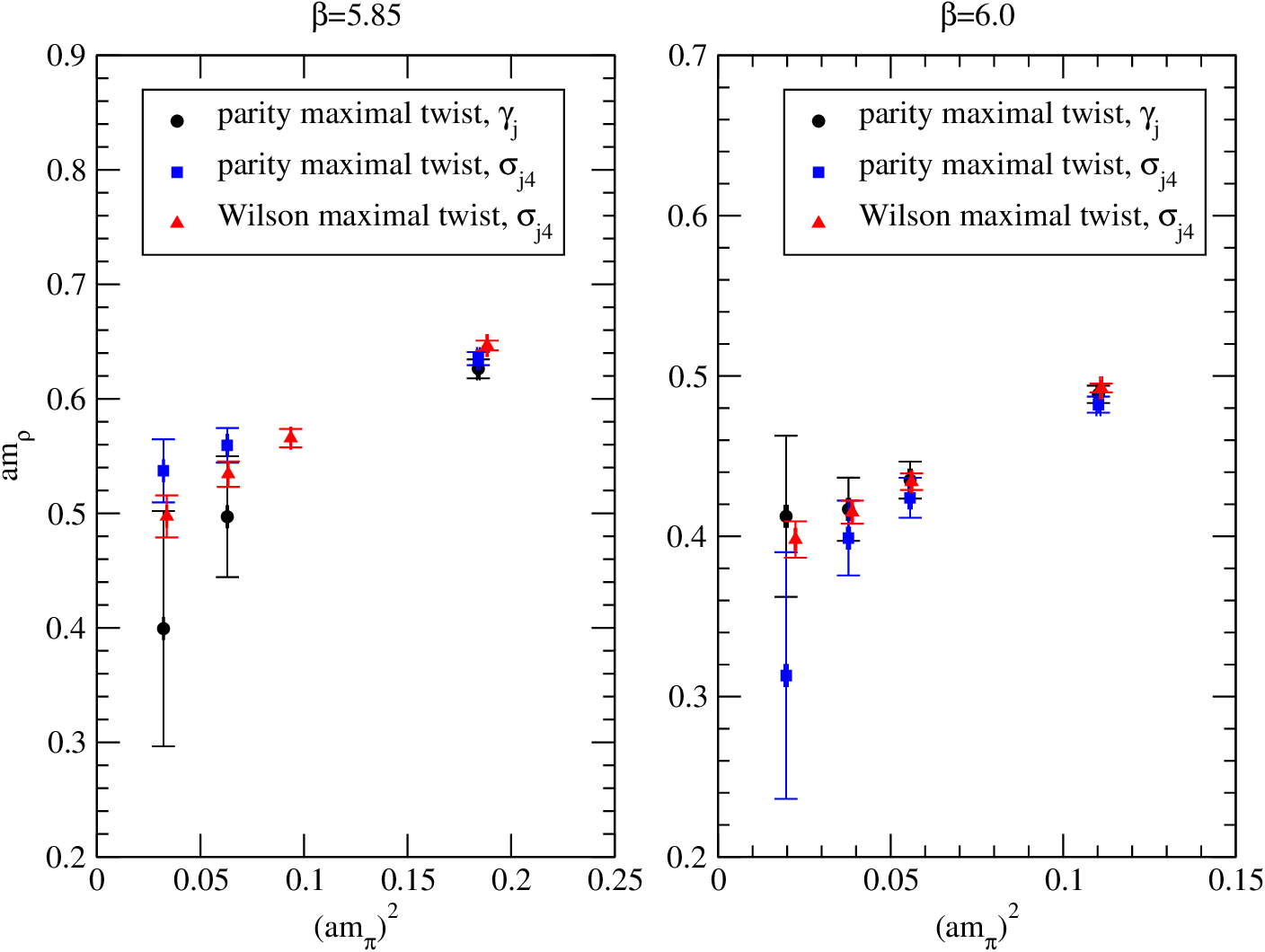}
\caption{Vector meson mass as a function of the pseudoscalar meson mass
         squared, calculated using the parity conservation definition of maximal
         twist: $\omega=\pi/2$ in Eq.~(\protect\ref{twistangle}).
         Data calculated using the Wilson $\kappa_{cW}$ definition, 
         Fig.~\protect\ref{fig:mpivsmuwilson},
         are shown for comparison.}
\label{fig:mrhovsmpiparity}
\end{figure}

\begin{figure}[tb]
\vspace{4mm}
\includegraphics[width=10cm]{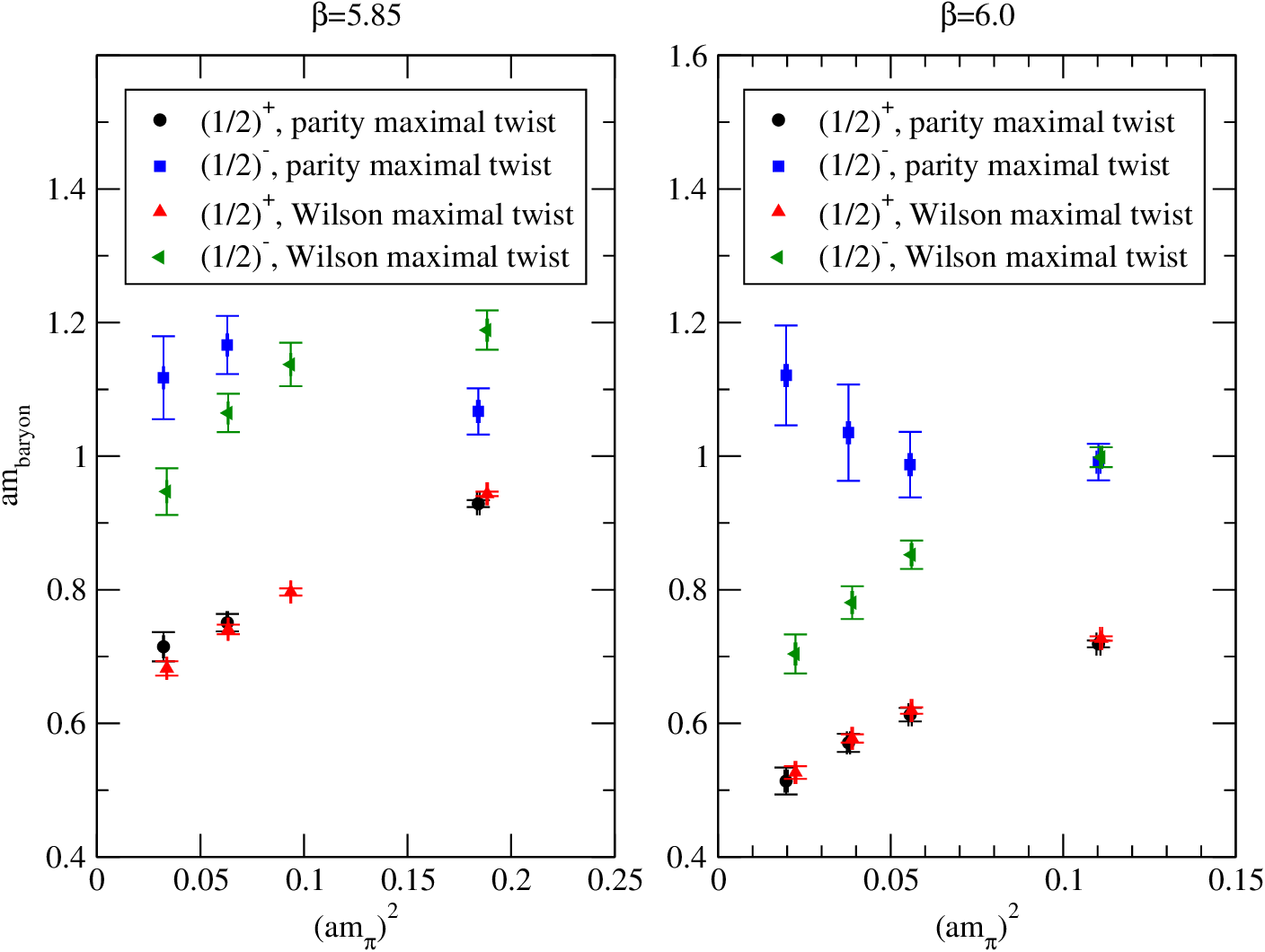}
\caption{Spin 1/2 baryon masses (both parities) as functions
         of the pseudoscalar meson mass squared, calculated using the parity
         conservation definition of maximal
         twist: $\omega=\pi/2$ in Eq.~(\protect\ref{twistangle}).
         Data calculated using the Wilson $\kappa_{cW}$ definition, 
         Fig.~\protect\ref{fig:mpivsmuwilson},
         are shown for comparison.}
\label{fig:mNvsmpiparity}
\end{figure}

\begin{figure}[tb]
\vspace{4mm}
\includegraphics[width=10cm]{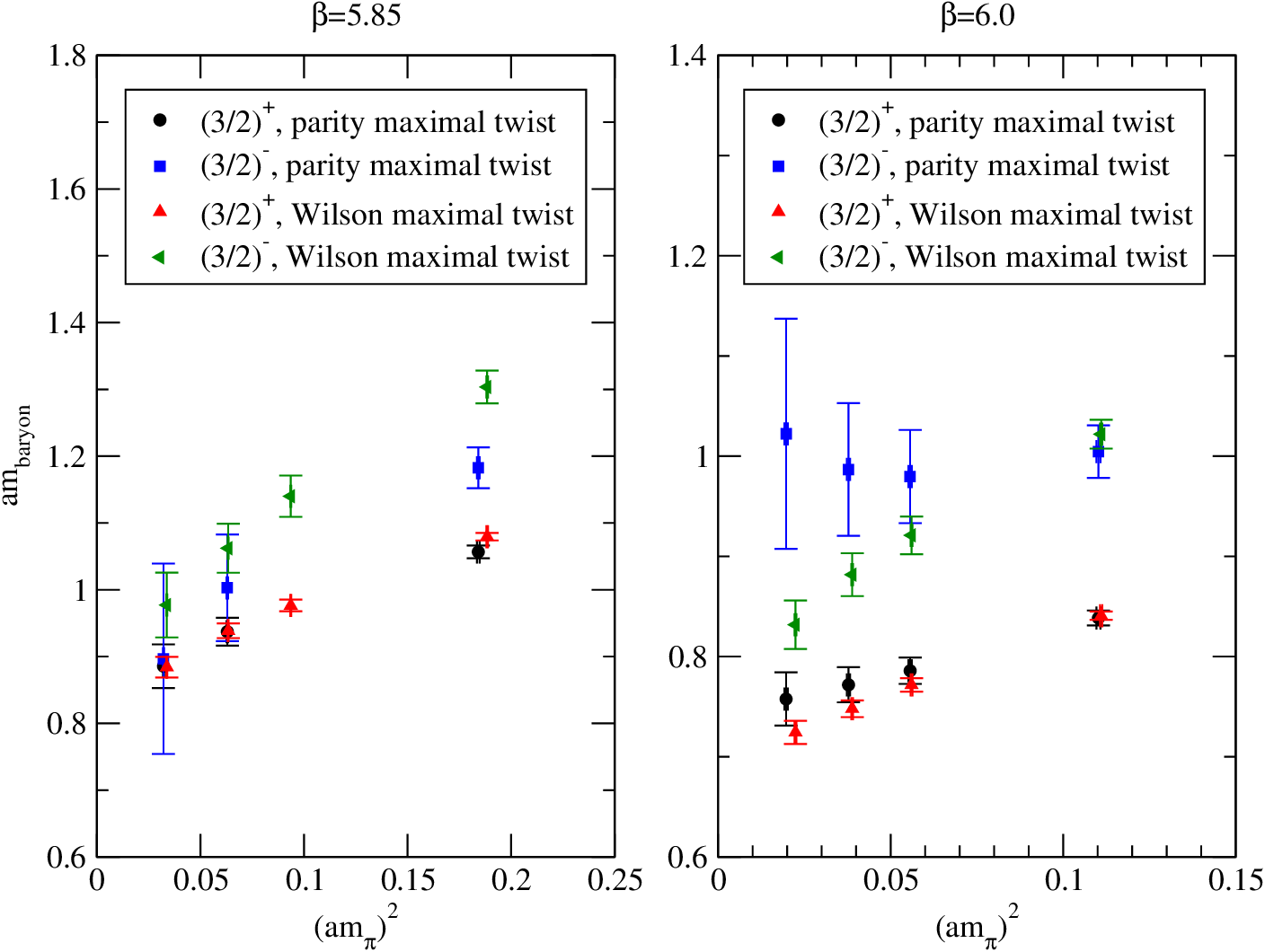}
\caption{Spin 3/2 baryon masses (both parities) as functions
         of the pseudoscalar meson mass squared, calculated using the parity
         conservation definition of maximal
         twist: $\omega=\pi/2$ in Eq.~(\protect\ref{twistangle}).
         These data are averages of the $\Delta^{++}$ and $\Delta^-$ masses.
         Data calculated using the Wilson $\kappa_{cW}$ definition, 
         Fig.~\protect\ref{fig:mpivsmuwilson},
         are shown for comparison.}
\label{fig:mDvsmpiparity}
\end{figure}

\begin{figure}[tb]
\vspace{4mm}
\includegraphics[width=10cm]{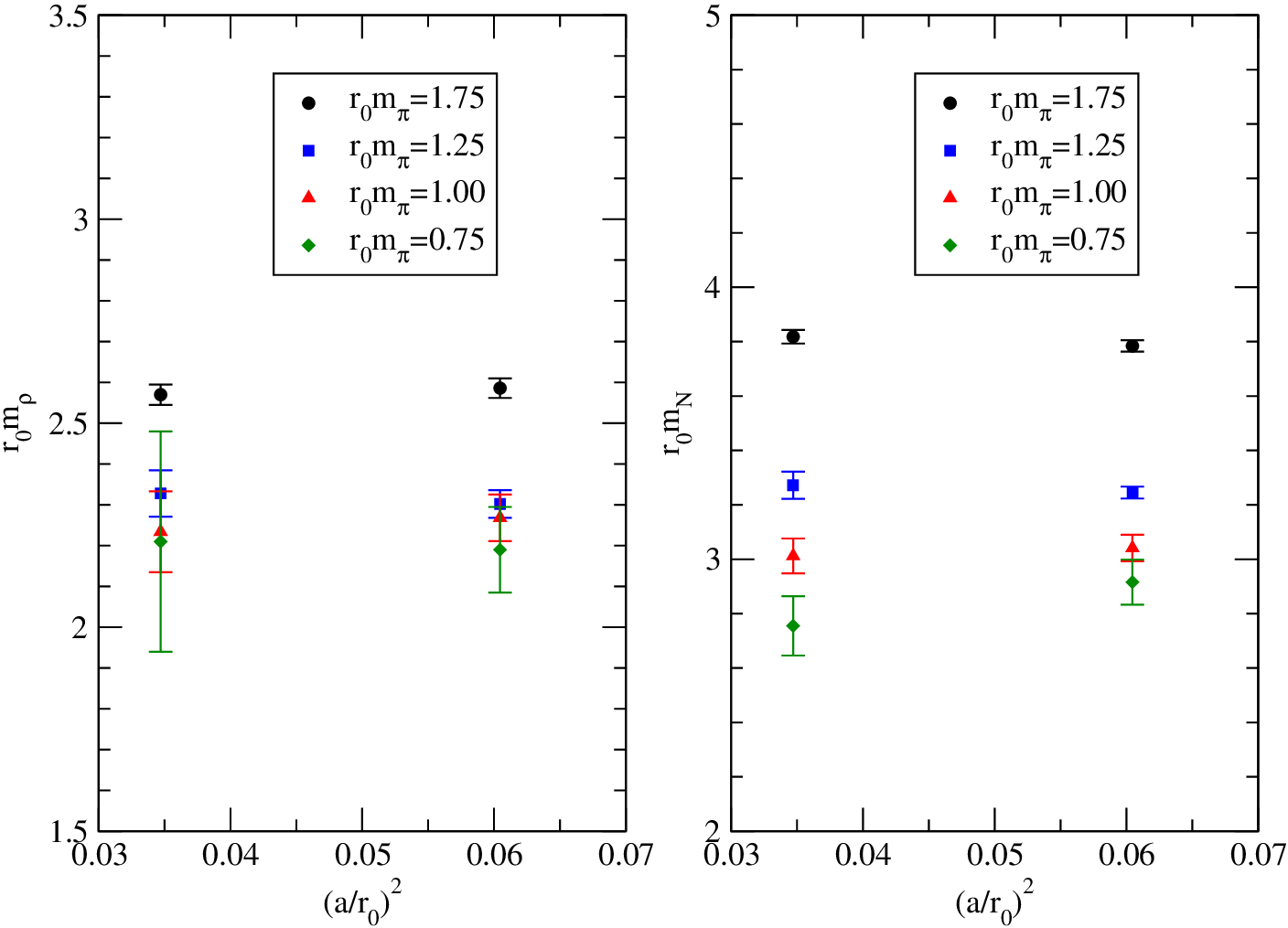}
\caption{Vector meson and $(1/2)^+$ baryon masses as functions
         of lattice spacing squared, calculated using the parity
         definition of maximal
         twist: $\omega=\pi/2$ in Eq.~(\protect\ref{twistangle}).
         Axes are in units of $r_0=0.5$ fm.}
\label{fig:scalingparity}
\end{figure}

Beyond the pseudoscalar meson, there is very little difference in the
computed spectrum between the two definitions of maximal twist.
The $\rho$ meson, spin-1/2 baryons and spin-3/2 baryons are displayed in
Figs.~\ref{fig:mrhovsmpiparity}, \ref{fig:mNvsmpiparity} and
\ref{fig:mDvsmpiparity}
respectively.  Clear signals are obtained for the $\rho$, nucleon and $\Delta$
hadrons, but data for the negative parity states do not show a convincing
signal.  The $\rho$ meson and nucleon masses can be 
seen to be independent of lattice spacing in Fig.~\ref{fig:scalingparity}
over the whole range of quark masses that were investigated.

\section{Flavor breaking and the $\Delta(1232)$ multiplet}\label{sec:flavor}

For the Wilson lattice action ($\mu=0$) with degenerate quark masses, 
the up and down quark propagators
are identical so hadron masses within any isospin multiplet are degenerate.
In tmLQCD, the up and down propagators (let's name them $U(x,y)$ and $D(x,y)$)
differ by having opposite signs for the $\mu$ term, so conservation of isospin
is not guaranteed.  However, tmLQCD does satisfy
\begin{eqnarray}
U(x,y) &=& \gamma_5 D^\dagger(y,x) \gamma_5, \\
D(x,y) &=& \gamma_5 U^\dagger(y,x) \gamma_5.
\end{eqnarray}
This leads to the equality of any given hadron two-point correlator with its
up-down interchanged hermitian conjugate.
Since masses are obtained from the real parts of correlators, we arrive at the
following degeneracies:
\begin{center}
\begin{tabular}{c}
$\pi^+$ and $\pi^-$, \\
$\rho^+$ and $\rho^-$, \\
proton and neutron, \\
$\Delta^{++}$ and $\Delta^-$, \\
$\Delta^+$ and $\Delta^0$.
\end{tabular}
\end{center}
The difference between charged and neutral pion masses offers a measure of
isospin violation, though it would require the computation of disconnected
quark diagrams.  Here, we consider a less expensive alternative:
$m_{\Delta^{++,-}}-m_{\Delta^{+,0}}$.

\begin{figure}[tb]
\vspace{4mm}
\includegraphics[width=10cm]{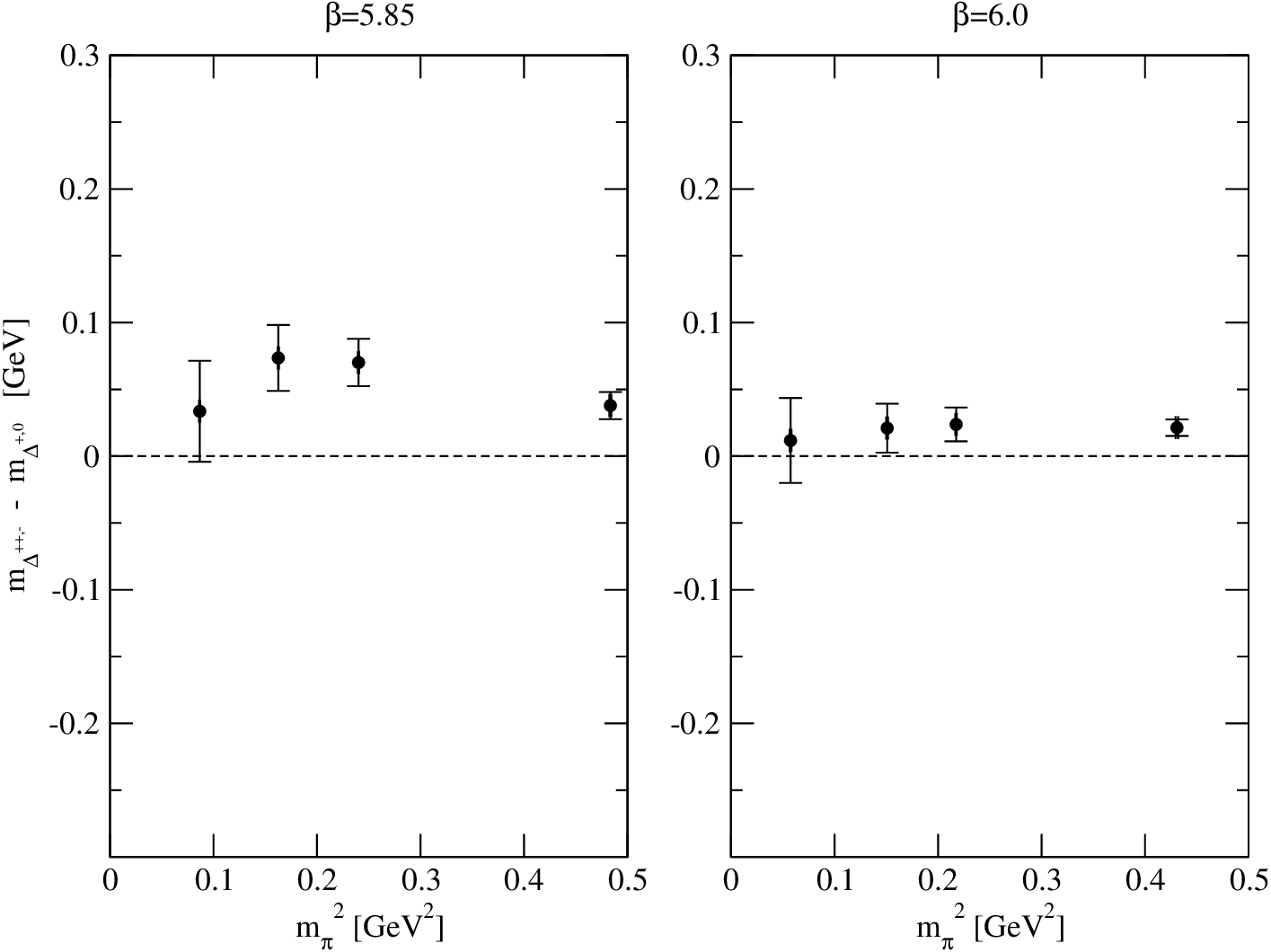}
\caption{Flavor splitting within the $\Delta(1232)$ multiplet as a function
         of the pseudoscalar meson mass squared.
         The standard mass parameter was held fixed to the value obtained from the
         Wilson $\kappa_{cW}$ definition of maximal twist, 
         Eq.~(\protect\ref{kappacWdef}).}
\label{fig:flavorwilson}
\end{figure}

\begin{figure}[tb]
\vspace{4mm}
\includegraphics[width=10cm]{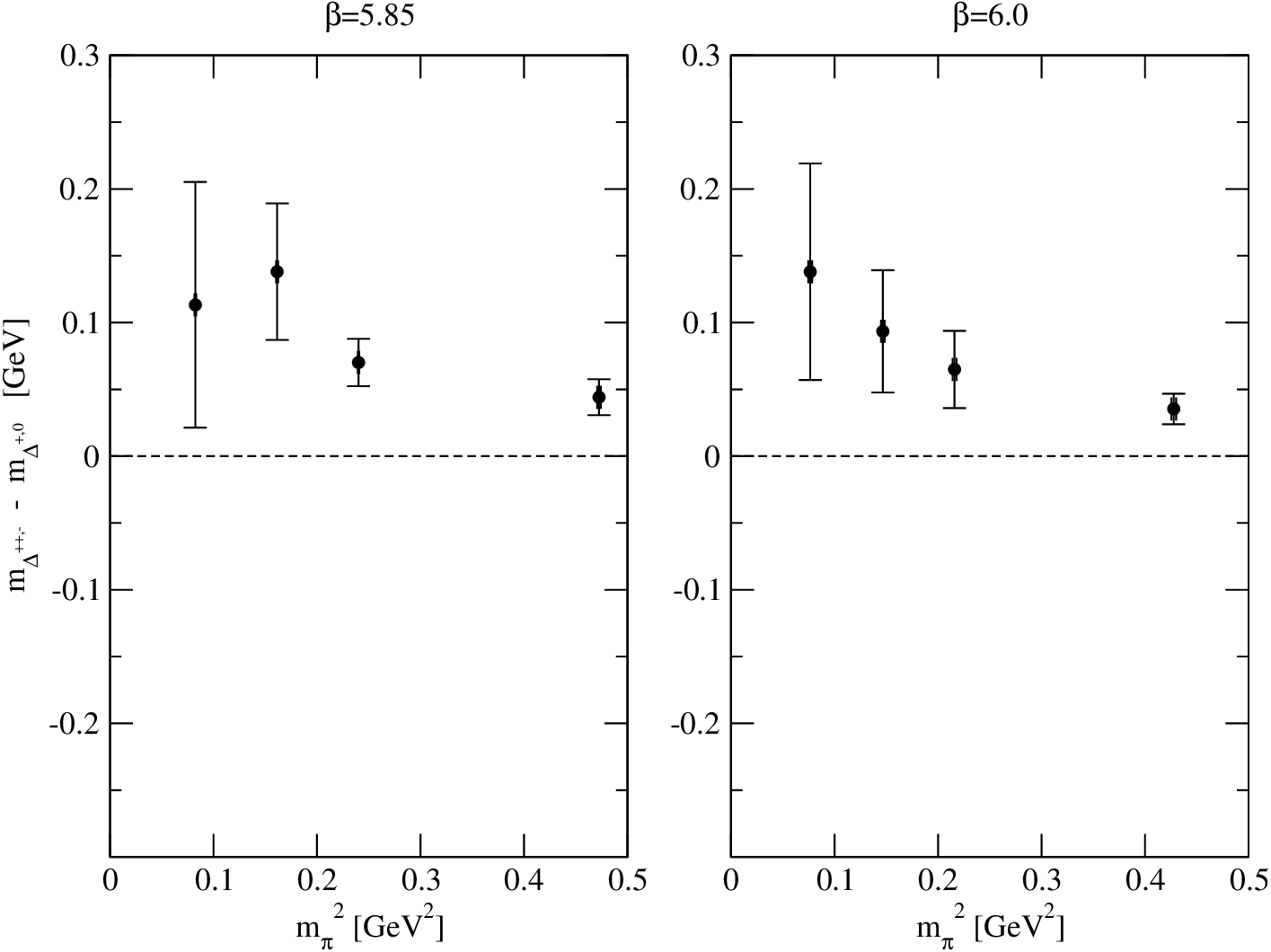}
\caption{Flavor splitting within the $\Delta(1232)$ multiplet as a function
         of the pseudoscalar meson mass squared, calculated using the parity
         conservation definition of maximal
         twist: $\omega=\pi/2$ in Eq.~(\protect\ref{twistangle}).}
\label{fig:flavorparity}
\end{figure}

Figs.~\ref{fig:flavorwilson} and \ref{fig:flavorparity} show our results for
this $\Delta$ splitting in physical units, as obtained respectively from
Wilson $\kappa_{cW}$
maximal twist and parity maximal twist.  No significant difference
is found
between the two definitions of maximal twist for this observable.  Based on
the better statistics of Fig.~\ref{fig:flavorwilson} (1000 configurations
rather than only 300), we see evidence that
$m_{\Delta^{++,-}}-m_{\Delta^{+,0}}$ decreases with decreasing
lattice spacing, as expected.

\section{Summary}\label{sec:outlook}

Twisted mass lattice QCD offers significant advantages over the usual
Wilson or clover\cite{clover} actions.
First ``exceptional configurations'' are 
eliminated and secondly there is the possibility, by tuning the mass 
parameter, of removing $O(a)$ lattice spacing errors automatically.
Our results add to the evidence that these advantages are realized in 
numerical simulations.
The spectrum of hadron masses was computed in tmLQCD for
two lattice spacings and two definitions of maximal twist
(where the spectrum is expected to be automatically $O(a)$ improved).
Four values of the quark mass were used, the lightest corresponding to a
pseudoscalar-meson to vector-meson mass ratio of about $1/3$. No problems
with convergence of the quark propagator calculation were observed so 
the quark mass could be pushed even lower.

There are different ways to define the twist angle and hence determine
maximal twist.
With the parity conservation definition of maximal twist,
Eq.~(\ref{twistangle}), the pion decay constant was found to be independent of
lattice spacing for all quark masses, though this was not true for the
definition of maximal twist that relies on the Wilson critical hopping
parameter.  No significant lattice spacing dependence
in vector meson or nucleon masses was observed over the whole range of
quark mass for both definitions of maximal twist.
Additional calculations at more lattice spacings
are needed before definitive statements about scaling can be made. 
However, it is already clear that maximally-twisted tmLQCD is 
competitive with other improved actions and, of course, much better than the 
unimproved Wilson action.

Phenomenologically reasonable results were obtained for the pion, $\rho$ meson,
nucleon and $\Delta(1232)$. The tensor operator was found to be useful for 
attaining the $\rho$ mass with a smaller uncertainty. Hints of the negative 
parity baryons were also obtained. To do better with the baryon resonances 
probably requires a more sophisticated treatment with operators optimized 
for particular states \cite{Basak}. 

For a fixed value of the twisted mass parameter $\mu$,
the parity conservation definition of maximal twist
produced a smaller pion mass than the Wilson 
$\kappa_{cW}$ definition.
Since the pion mass squared should be minimized at maximal twist,
this suggests that the parity conservation definition is a better definition, 
at least for this observable.
In addition, only the parity conservation definition led to a
realistic extrapolation for the pion decay constant.
Note that implementing this definition required tuning of
the standard mass parameter for every twisted mass value which adds to the
overall cost of doing the calculations.
For the rest of the mass spectrum, our data do not show a statistically
significant difference between the two definitions of maximal twist. 

A lattice artifact of tmLQCD is flavor symmetry breaking. This can manifest
itself as a mass difference between charged and neutral pions. However, the 
calculation of the neutral pion mass in the presence of flavor symmetry 
breaking is a formidable task since it involves disconnected quark line
contributions. In this work we considered mass splittings in the isospin-3/2
$\Delta(1232)$ multiplet which involved no disconnected quark line 
contribution. From our higher statistics run there is evidence that
the flavor symmetry breaking effect is present and that it decreases at 
smaller lattice spacing.

\acknowledgments

This work was supported in part by
the Natural Sciences and Engineering Research Council of Canada,
the Canada Foundation for Innovation,
the Canada Research Chairs Program
and the Government of Saskatchewan.

\newpage

\begin{table}[tb]
\caption{Parameters used in the simulations.  Parameters in parenthesis
denote the one instance where two definitions of maximal twist happen to
coincide so new data were not generated.}
\label{tab:parameters}
\begin{tabular}{ccccl}
~~~$\beta$~~~ & ~~~\#sites~~~ & \#configurations & ~~~~~~$aM$~~~~~~ 
& ~~~$a\mu$~~~ \\
\hline
5.85 & 20$^3\times40$ & 1000 & -0.9071 & 0.0376 \\
     &                &      & -0.9071 & 0.0188 \\
     &                &      & -0.9071 & 0.01252 \\
     &                &      & -0.9071 & 0.00627 \\
\cline{3-5}
     &                &  300 & -0.8965 & 0.0376 \\
     &                &      & (-0.9071) & (0.0188) \\
     &                &      & -0.9110 & 0.01252 \\
     &                &      & -0.9150 & 0.00627 \\
\hline
6.0  & 20$^3\times48$ & 1000 & -0.8135 & 0.030 \\
     &                &      & -0.8135 & 0.015 \\
     &                &      & -0.8135 & 0.010 \\
     &                &      & -0.8135 & 0.005 \\
\cline{3-5}
     &                &  300 & -0.8110 & 0.030 \\
     &                &      & -0.8170 & 0.015 \\
     &                &      & -0.8195 & 0.010 \\
     &                &      & -0.8210 & 0.005 \\
\end{tabular}
\end{table}

\begin{table}[tb]
\caption{The mass spectrum and pion decay constant from our simulations.
The vector meson masses are obtained from $\bar\psi\sigma_{j4}\tau^\pm\psi$ or
$\bar\psi\gamma_j\tau^\pm\psi$, whichever has the smaller uncertainty.
Spin 3/2 masses are given
as averages over pairs within the multiplet.}
\label{tab:masses}
\begin{tabular}{ccccccccc}
$\beta$ & $aM$ & $a\mu$ & $am_\pi$ & $af_\pi$ & $am_\rho$ & $am_N$
 & $am_{\Delta^{++},\Delta^-}$ & $am_{\Delta^+,\Delta^0}$ \\
\hline
5.85 & -0.9071 & 0.0376 & 0.4341(4) & 0.1144(7) & 0.647(4) & 0.943(3) & 1.079(6) & 1.056(6) \\
     & -0.9071 & 0.0188 & 0.3060(4) & 0.0995(7) & 0.566(8) & 0.797(5) & 0.976(9) & 0.933(11) \\
     & -0.9071 & 0.01252 & 0.2518(5) & 0.0923(7) & 0.534(11) & 0.741(7) & 0.939(11) & 0.893(16) \\
     & -0.9071 & 0.00627 & 0.1837(6) & 0.0812(7) & 0.497(18) & 0.682(11) & 0.884(15) & 0.863(22) \\
     & -0.8965 & 0.0376 & 0.4292(8) & 0.1123(10) & 0.635(6) & 0.929(5) & 1.057(5) & 1.029(10) \\
     & -0.9110 & 0.01252 & 0.2509(8) & 0.0933(10) & 0.559(15) & 0.751(13) & 0.937(21) & 0.851(30) \\
     & -0.9150 & 0.00627 & 0.1793(10) & 0.0876(11) & 0.54(3) & 0.715(22) & 0.886(33) & 0.815(54) \\
\hline
6.0  & -0.8135 & 0.030 & 0.3332(5) & 0.0859(5) & 0.493(3) & 0.727(3) & 0.841(4) & 0.830(5) \\
     & -0.8135 & 0.015 & 0.2367(6) & 0.0742(5) & 0.434(5) & 0.619(5) & 0.772(7) & 0.760(8) \\
     & -0.8135 & 0.010 & 0.1971(6) & 0.0679(5) & 0.415(7) & 0.577(6) & 0.748(8) & 0.737(10) \\
     & -0.8135 & 0.005 & 0.1497(8) & 0.0563(5) & 0.398(11) & 0.526(9) & 0.724(12) & 0.718(18) \\
     & -0.8110 & 0.030 & 0.3320(9) & 0.0855(9) & 0.482(5) & 0.719(5) & 0.839(7) & 0.821(7) \\
     & -0.8170 & 0.015 & 0.2359(11) & 0.0759(10) & 0.424(12) & 0.613(10) & 0.786(13) & 0.753(14) \\
     & -0.8195 & 0.010 & 0.1944(12) & 0.0723(10) & 0.417(20) & 0.571(14) & 0.772(18) & 0.724(20) \\
     & -0.8210 & 0.005 & 0.1404(17) & 0.0673(11) & 0.412(50) & 0.514(20) & 0.758(27) & 0.688(37) \\
\hline
\end{tabular}
\end{table}

% figure 1

% figure 2

% figure 3

% figure 4

% figure 5

% figure 6

% figure 7

% figure 8

% figure 9

% figure 10

% figure 11

% figure 12

% figure 13

% figure 14

% figure 15

% figure 16

\end{document}